\documentclass[twocolumn,trackchanges]{aastex7}

\usepackage{bm} 
\usepackage{soul}

\begin{document}

\title{Probing small-scale dark matter clumping with the large-scale 21-cm power spectrum}

\author[orcid=0000-0001-6129-0118,sname='Sikder']{Sudipta Sikder}
\affiliation{School of Physics and Astronomy, Tel-Aviv University, Tel-Aviv, 69978, Israel}
\email[show]{sudiptas@mail.tau.ac.il}

\author[orcid=0000-0002-7464-7857,gname=Hyunbae, sname='Park']{Hyunbae Park} 
\affiliation{Center for Computational Sciences, The University of Tsukuba, 1 Chome-1-1 Tennodai, Tsukuba, Ibaraki 305-8577, Japan}

\affiliation{Computational Cosmology Center, Lawrence Berkeley National Laboratory, 1 Cyclotron Road, Berkeley, California 94720, USA}

\email{parkhb@ccs.tsukuba.ac.jp}

\author[orcid=0000-0002-1557-693X,gname=Rennan, sname='Barkana']{Rennan Barkana} 
\affiliation{School of Physics and Astronomy, Tel-Aviv University, Tel-Aviv, 69978, Israel}
\email{barkana@tauex.tau.ac.il}

\author[orcid=0000-0001-7925-238X,gname=Naoki, sname='Yoshida']{Naoki Yoshida} 
\affiliation{Department of Physics, School of Science, The University of Tokyo, 7-3-1 Hongo, Bunkyo, Tokyo 113-0033, Japan}

\affiliation{Kavli Institute for the Physics and Mathematics of the Universe, 5-1-5 Kashiwanoha Kashiwa, Chiba 277-8583, Japan}

\affiliation{Max Planck Institut für Astophysik, Karl-Schwarzschild-Str. 1 Garching, D-85741, Germany}

\email{naoki.yoshida@ipmu.jp}

\author[orcid=0000-0002-1369-633X
,gname=Anastasia, sname='Fialkov']{Anastasia Fialkov} 
\affiliation{Institute of Astronomy, University of Cambridge, Madingley Road Cambridge, CB3 0HA, UK}

\affiliation{Kavli Institute for Cosmology, Madingley Road Cambridge, CB3 0HA, UK}

\email{anastasia.fialkov@gmail.com}

\begin{abstract}

The 21-cm line of hydrogen is the most promising probe of the Dark Ages and Cosmic Dawn. We combine hydrodynamical simulations with a large-scale grid in order to calculate the effect of non-linear structure formation on the large-scale 21-cm power spectrum, focusing on redshifts $z=20-40$. As the clumping effect arises from small-scale density fluctuations, it offers a unique opportunity to probe the standard cold dark matter model in a new regime and thus potentially investigate the properties of dark matter. To this end, we also study a warm dark matter $-$ like model with a Gaussian cutoff on a scale of 50~kpc. We find that clumping has a significant impact on the large-scale 21-cm power spectrum, requiring a substantial correction to standard theoretical predictions. For example, for the Dark Ages case at $z=30$ and wavenumber $k=0.05$~Mpc$^{-1}$, small-scale clustering enhances the 21-cm power spectrum by 13\%. Once Lyman-$\alpha$ coupling kicks in due to the first stars, the 21-cm signal strengthens, and the effect of clumping grows; it suppresses the observable power spectrum at $z=20$ by 45\%, while the warm dark matter $-$ like model has less than half the clumping impact. The clumping effect is significantly higher than the sensitivity of the planned Square Kilometre Array (SKA) AA$^\star$ configuration, by up to a factor of 20 for standard cold dark matter, though detection will require separation from foregrounds and from astrophysical contributions to the 21-cm power spectrum.

\end{abstract}



\section{Introduction}

The standard cosmological model is primarily based on the foundation of the hot Big Bang. For the first 380,000 years after the Big Bang, the Universe existed as an opaque plasma of particles and photons, with free electrons constantly scattering photons. As the Universe expanded and cooled, protons and electrons recombined to form neutral hydrogen, making the Universe transparent to radiation. At this juncture, photons decoupled from matter and began streaming freely through space. These photons, redshifted due to cosmic expansion, are detectable today as the cosmic microwave background (CMB). Primordial density fluctuations, observed through experiments such as the Planck satellite \citep{planckcollaboration18}, were amplified by gravity over time, giving rise to the large-scale structures visible in the Universe today. Complementing these early Universe observations, galaxy surveys probe a much later cosmos, capturing light from stars and galaxies \citep{Carniani2024, Helton2025} that span redshifts from $z\sim 14$ (corresponding to a cosmic age of 300 million years) to the present day ($z=0$). 

Between the CMB and the local Universe lies a vast, largely unexplored era --- from the cosmic Dark Ages through Cosmic Dawn to the Epoch of Reionization (EoR). During the Dark Ages, before the first stars formed, the Universe was nearly homogeneous and driven by well-understood physical processes. Cosmic Dawn marked the ignition of the first stars and galaxies, while the EoR saw the Universe transitioning from neutral atoms to ionized plasma. Spanning redshifts $z \approx 200 - 6$, these eras witnessed dramatic evolution driven by gravitational collapse, star formation, and associated astrophysical processes. The 21-cm signal, arising from the hyperfine transition of neutral hydrogen (HI), is the most promising probe for studying these transformative phases \citep{furlanetto06,barkana18book,Mesingerbook}. This signal is highly sensitive to both astrophysical processes, such as star formation and radiative feedback, and the underlying dark matter density field, making it an invaluable tool for testing the standard Cold Dark Matter (CDM) model and exploring alternative scenarios such as Warm Dark Matter (WDM). Since the large-scale distribution and behavior of dark matter have been well established observationally, for further progress we must focus on small scales. Linear theory then becomes inadequate to capture structure formation due to the density contrast significantly exceeding unity. High-resolution hydrodynamical simulations are essential in this regime, as they accurately model and follow the complex interplay of gravitational collapse, gas dynamics, non-equilibirum chemistry, and radiative processes, providing a detailed picture of how small-scale density fluctuations and dark matter properties shape structure formation across cosmic epochs. 

To detect the cosmological 21-cm signal and leverage its sensitivity to various cosmological and astrophysical processes, two primary observational strategies are employed. The first, a simpler method, measures the sky-averaged radio intensity as a function of frequency (and thus redshift) using a dipole antenna. Current efforts to measure this global 21-cm signal include REACH \citep{de_lora2022}, MIST \citep{mist2024}, RHINO \citep{bull2024}, PRIZM \citep{prizm2019}, and SCI-HI \citep{sci-hi2014}; the two ongoing experiments that have achieved the most notable results so far are EDGES with a tentative detection \citep{bowman18}, and SARAS~3 with its conflicting evidence that challenges that detection \citep{SARAS3}. The second approach uses a radio interferometer to measure spatial fluctuations in the 21-cm signal, currently focused on a statistical detection through the 21-cm power spectrum. Ongoing projects include the MWA \citep{Trott:2020}, LEDA \citep{leda_garsden}, and NenuFar \citep{Munshi2024}; the ongoing projects that have provided the most stringent upper limits on the 21-cm power spectrum are HERA \citep{Abdurashidova_2023} and LOFAR \citep{LOFAR-EoR:2020}. Great promise lies in the upcoming Square Kilometre Array (SKA) \citep{koopmans15}, with construction underway and scientific data expected in 5 years.

Observations of the 21-cm signal are expected to be constrained to relatively large scales. The global 21-cm signal effectively measures the cosmically averaged 21-cm brightness temperature at each redshift, so it obviously smooths over small scales. Meanwhile, the 21-cm power spectrum should be limited to Mpc scales and larger since a higher resolution would require enormous collecting areas in order to maintain enough sensitivity to detect the weak cosmic signal. It might be expected that these observations of the 21-cm signal, that are smoothed over small scales, can only probe physics that occurs on large scales, but this is not the case due to the strong non-linearity of the 21-cm signal. This is a major difference from the CMB, which is far more linear due to the much higher redshift ($\sim$1100) plus the fact that Silk damping \citep{1968ApJ...151..459S} wipes out small-scale fluctuations. In 21-cm cosmology, small-scale fluctuations do not average out but can leave a clear signature even in observations that are smoothed on large scales.

Numerical simulations indicate that non-linear clumping enhances the global 21-cm absorption signal in the late Dark Ages \citep{Ahn2006, Shapiro2006} and, with saturated Lyman-$\alpha$ coupling assumed, diminishes it during Cosmic Dawn \citep{Xu2018, Xu2021}. However, these simulations, limited to box sizes of 0.7~Mpc \citep{Ahn2006, Shapiro2006} or 8 $h^{-1}$ Mpc \citep{Xu2021}, neglected large-scale density fluctuations as well as the effects of the baryon - dark-matter streaming velocity \citep{tseliakhovich10}. This streaming velocity affects the formation of the first galaxies in
a non-uniform manner by delaying the growth of baryonic clumps, which can introduce 21-cm fluctuations at $\sim$100 Mpc scales during Cosmic Dawn \citep[e.g.,][]{visbal12,2019PhRvD.100f3538M}. This should similarly affect all
nonlinear density clumps. Analytical models suggest that the streaming velocity induces large-scale 21-cm fluctuations during the Dark Ages \citep{2014alihamoud}, while numerical simulations with saturated coupling found a similar effect during the EoR \citep{Cain2020}. However, both of these predicted effects are quite small, and well below upcoming observational capabilities.

Recently \citep{Park} we combined hydrodynamical simulations with a large-scale grid in order to precisely calculate the effect of nonlinear structure formation on the global (sky-averaged) 21-cm radio intensity. We showed that clumping effects are significant enough to be detectable in the global signal. This can in principle be detected
cleanly and unambiguously during the Dark Ages, but that weak signal (with a maximum amplitude of half a milli-Kelvin at $z=27$) requires an array of global signal antennae. During Cosmic Dawn, when stellar radiation boosts the signal, clumping can change the signal by as much as 15~mK at $z=20$, and a single global antenna suffices for detection, but the clumping effect
must then be modeled and separated from the effect of the stars. The clumping effect offers a unique window into the strength of primordial density fluctuations on small scales (corresponding to the mass scale of the smallest dwarf galaxies), enhancing our ability to constrain the nature of dark matter as well as non-standard cosmological models.
The ability to use observations at early cosmic times, before the complexities of non-linear astrophysics had time to substantially modify the Universe, is a further advantage. The direct clumping effect arises from the non-linear, filamentary cosmic web, and not primarily from virialized objects. An indirect effect arises from the Ly$\alpha$ coupling, which depends on the star formation rate in the first galaxies embedded in the cosmic web, and has been suggested as a way to constrain small-scale matter clumping \citep{2020PhRvD.101f3526M}, albeit with large uncertainties due to our limited understanding of the astrophysics of star formation and stellar feedback.

In this work, we predict for the first time a significant, potentially observable  effect of the non-linear cosmic web of intergalactic hydrogen on the 21-cm power spectrum, during the Dark Ages and Cosmic Dawn. Our multifaceted approach of combining small-scale hydrodynamical simulations with a large-scale semi-numerical grid allows us to bridge the enormous range of scales involved. Our goal is to elucidate how small-scale clumping, modulated by large-scale density and velocity fluctuations, influences the spatial variations of the 21-cm signal and can potentially be used to constrain the properties of dark matter.

\section{Methods}

\subsection{Numerical simulation}

\label{s:sim}

Our method is based on combining hydrodynamical simulations with a large-scale grid \citep{Park}, in order to span an extremely wide range of scales. For the small-scale simulations, we use a modified version of the GADGET package \citep{2001NewA....6...79S,2005MNRAS.364.1105S,2006ApJ...652....6Y,2007ApJ...663..687Y}, which includes chemical reactions involving the species of primordial hydrogen and helium, along with Compton heating from interactions between free electrons and CMB photons. We employ the BaryonCDM Cosmological Initial Condition (IC) Generator for Small Scales (BCCOMICS) \citep{Ahn2016, Ahn2018} to initialize simulations at redshift $z_{i}=200$. Since at sub-megaparsec scales, the relative streaming velocity between baryons and dark matter significantly influences baryonic structure formation during the Dark Ages, accurate simulation of this streaming velocity effect is essential. BCCOMICS independently solves structure growth equations for dark matter and baryons, incorporating the streaming velocity to generate density and velocity fields at our initial redshift $z_{i}=200$. It also allows for non-zero initial mean overdensities in the simulation box, enabling the modeling of structure formation across diverse environments. 

The simulation is initialized in a 3 Mpc cubic volume with $512^3$ dark matter particles ($m_{\rm{DM}}=6670 \ M_{\odot}$) and $512^3$ gas particles ($m_{\rm{gas}}=1250 \ M_{\odot}$) for Smoothed Particle Hydrodynamics (SPH) calculations, to model the structure formation at sub-Mpc scales. The fixed mass resolution of SPH effectively models gas physics in dense regions, though it struggles with abrupt discontinuities from strong shocks. Such shocks are rare during the Dark Ages and early Cosmic Dawn, and mild shocks from streaming velocities align well with results from Lagrangian and Eulerian simulations. To focus on the gas temperature ($T_{\rm{gas}}$) without star formation, cooling by hydrogen atoms and molecules is disabled to prevent gas collapse in minihalos (masses $\approx 10^6 M_\odot$, starting at $z\sim 30$). Feedback from star formation, even after the first stars form, affects only a small fraction of baryons, making $T_{\rm{gas}}$ statistics reliable down to early Cosmic Dawn ($z \approx 20$). 

In broad surveys covering a wide range of possible astrophysical parameters of high-redshift galaxies, we showed that generically the first substantial astrophysical 21-cm effect is Lyman-$\alpha$ coupling \citep{2017MNRAS.472.1915C, 2018MNRAS.478.2193C, 2021MNRAS.506.5479R}. The earliest cosmic heating, though highly uncertain, likely comes later, with cosmic reionization coming later still. Thus, we expect an early stage of Cosmic Dawn during which the only important astrophysical effect is Lyman-$\alpha$ coupling from stellar radiation. Lyman-$\alpha$ radiation travels large distances and thus is dominated by sources at distances that are much larger than our simulation box size. In this work we do not explicitly include star formation processes in our simulations; we only aim to illustrate the strength of the gas clumping effect in the presence of significant Lyman-$\alpha$ coupling, using two levels of coupling as explained below. 

The initial baryon-dark matter streaming velocity $V_{\rm bc,i}$ at cosmic recombination ($z \sim 1100$) follows a Boltzmann distribution, with a standard deviation $V_{\star} = 28$ km/s \citep{tseliakhovich10}. We run simulation boxes with initial conditions for 17 combinations of streaming velocity and initial overdensity $\bar\delta$ of the simulation box. Fifteen cases are all the combinations of $V_{\rm bc,i}=0, V_{\star}, 2 V_{\star}$ and box-averaged overdensity $\bar\delta/\sigma_{\star}=-2, -1, 0, 1, 2$, where $\sigma_{\star}=0.014$ (the standard deviation of density in 3~Mpc boxes at $z=200$). Two additional cases use $\delta/\sigma_{\star}=-3, 3$ with $V_{bc,i}=V_{\star}$. We note that the grid used in our small-scale simulations is relatively coarse. We have checked that the dependence on the parameters is smooth (often close to linear), and that adding the $3\sigma_{\star}$ points had a negligible effect on our overall results (below 1\%). Further convergence tests along with quantification of interpolation errors remain avenues for future investigation. For non-zero overdensity cases, the separate universe prescription \citep{Park2020, Ahn2018, Sirko2005} is employed to evolve structures, adjusting the comoving box size to reflect locally slower (overdense) or faster (underdense) cosmic expansion. We account for the Lagrangian to Eulerian conversion when calculating the mean 21-cm brightness of the box (since SPH is inherently a Lagrangian code), and include redshift space distortions (see \citet{Park} for further details). We assume standard cosmological parameters \citep{planckcollaboration18}.

In order to separate out the effect of clumping, in the results below we include a case of large-scale fluctuations only, which excludes small-scale clumping. To this end, we also run simulations boxes with uniform density (which varies with the same seven values listed above, reflecting large-scale fluctuations). Note that in this case the streaming velocity has no effect; it influences the evolution of dark matter and baryons only when there are density gradients, in which case the relative motions of density clumps change the distribution of gravitational forces.

\subsection{Warm dark matter model}

In the CDM model, fluctuations in the dark matter density field are expected to exist at small scales (high wavenumbers). This predicts numerous low-mass dark matter halos ($10^6 - 10^8 \ M_{\odot}$) that should host satellite galaxies. However, observations show fewer satellites, suggesting either suppressed star formation in these halos or an overprediction of substructure by CDM. Small-scale structure presents an opportunity to probe CDM in a new regime. This is an exciting prospect since various dark matter models beyond standard cold dark matter (CDM) produce small-scale cutoffs in the density power spectrum, including WDM (in which the dark matter has a significant initial velocity dispersion) \citep{WDM} and fuzzy dark matter (FDM, in which an extremely low dark matter particle mass makes quantum effects appear on galactic scales) \citep{fuzzy}. 

Current robust, direct observational constraints on the small-scale power spectrum extend up to comoving wavenumbers of $k\sim 10 \ \rm{Mpc}^{-1}$, primarily from Lyman-$\alpha$ forest measurements at $z=4-5.4$ \citep{2023PhRvD.108b3502V,2021PhRvL.126g1302R, 2017PhRvL.119c1302I}; these correspond to a minimum WDM mass of $\sim 3$~keV and a minimum FDM mass of $\sim 2\times 10^{-21}$~eV. However, these estimates are subject to systematic uncertainties related to reionization and the astrophysical heating of the intergalactic medium \citep{2021MNRAS.502.2356G}. An important advantage of 21-cm observations is that they probe high redshifts, when the Universe was more homogeneous, density fluctuations were more linear, and complex astrophysics had not yet had time to operate (in the Dark Ages, in particular). Also, the typical length scale probed by the clumping effect on the 21-cm signal is an order of magnitude smaller than current limits, corresponding to a mass scale that is 3 orders of magnitude smaller. 

In this work, to simulate WDM-like suppression, a set of 17 initial conditions is generated by applying Gaussian smoothing to the initial density and velocity  fields with a cutoff wavenumber $k_{\rm{cut}} = 100 \, h \, \rm{Mpc}^{-1}$ (where $h=0.6766$). We multiply each of the Fourier modes in the initial conditions by $\exp(-k^2/k^2_{\rm cut})$, to damp the fluctuation modes at wavenumber $k \gtrsim k_{\rm cut}$. Such a cutoff provides a useful generic model that can mimic the effects of various non-standard dark matter models, such as WDM or FDM (although these particular models produce additional dynamical effects that may somewhat modify the results); we refer to the cutoff case as the $100h$ WDM-like model. If we match the wavenumber at which the linear perturbation drops to half of the value in CDM, we find that this model is approximately equivalent to a 6.6~keV WDM mass and a $2.2\times 10^{-20}$~eV FDM mass. We choose the above value of $k_{\rm{cut}}$ for our illustrative example since it marks the characteristic scale above which arises about half of the clumping effect on the 21-cm global signal \citep{Park}, and we expect something similar with the 21-cm power spectrum. 

\subsection{Large-scale grid and the 21-cm power spectrum}

The strength of the 21-cm signal is expressed as the brightness temperature contrast between the spin temperature ($T_{\rm{S}}$) and the background radiation temperature ($T_{\rm{rad}}$) at the rest-frame 21-cm wavelength at redshift $z$ \citep{madau97, Pritchard2012, barkana18book}:
\begin{equation}
    T_{21} (\bm x, z) = \frac{T_{\rm S} (\bm x, z) - T_{\rm rad} (\bm x, z)}{1+z}(1 - e^{-\tau_{21}(\bm x, z)})\ , \label{eq:T21}
\end{equation}
where $\tau_{21}$ is the 21-cm optical depth. Usually (and in this paper) the background radiation is assumed to be the CMB at redshift $z$, so $T_{\rm rad} = T_{\rm CMB} = 2.725(1 + z)$ K, unless an excess radio background above the CMB is present \citep{feng18, ewall18, fialkov19, Reis2020, sikder2023}. The 21-cm signal is detectable against the CMB when $T_{\rm S}$ deviates from the CMB temperature ($T_{\rm CMB}$). Three key processes govern $T_{\rm S}$: scattering with CMB photons, atomic collisions, and Lyman-$\alpha$ coupling \citep{wouthuysen52, field58}. Their respective strengths are described by the  coupling coefficients, $x_{\rm rad}$, $x_{\rm c}$, and $x_{\alpha}$. From the equilibrium between these processes, $T_{\rm S}$ can be written as \citep{madau97}
\begin{equation}
T_{S}^{-1} = \frac{x_{\rm rad} T_{\rm rad}^{-1} + x_{\rm c} T_{\rm K}^{-1} + x_{\rm \alpha}T_{\rm C}^{-1}}{x_{\rm rad} + x_{\rm c} + x_{\rm \alpha}}\ , \label{eq:TS}
\end{equation}
where $T_{\rm K}$ is the kinetic temperature of the hydrogen atoms and $T_{\rm C}$ is the color temperature of the Lyman-$\alpha$ radiation field, which approaches $T_{\rm K}$ through repeated scattering.

There are two primary methods for measuring the 21-cm signal.
The first uses a single antenna to observe the sky and obtain the sky-averaged (global) spectrum, which traces the evolution of the cosmic mean signal with time or redshift.
The second method relies on interferometry to capture the evolution of spatial fluctuations in the 21-cm signal. This technique provides detailed spatial information across a broad range of wavenumbers at each redshift. In this work we focus on the power spectrum of the 21-cm brightness temperature fluctuations, defined by
\begin{equation}
\langle \tilde{\delta}_{T_{21}}(\bm{k})\tilde{\delta}_{T_{21}}^*(\bm{k^{\prime}}) \rangle = (2\pi)^3\delta_D(\bm{k}-\bm{k^{\prime}})P_{21}(k)\ ,
\end{equation}
where $\bm k$ is the comoving wavevector, $\delta_{D}$ is the Dirac delta function, and $\tilde{\delta}_{T_{21}}(\bm{k})$ is the Fourier transform of the 21-cm fluctuation $\delta_{T_{21}}(\bm{x})$, which is defined by $\delta_{T_{21}} (\bm{x}) = (T_{21}(\bm{x}) - \langle T_{21}\rangle)/\langle T_{21}\rangle$, with angular brackets denoting the average. We express the 21-cm power spectrum in terms of the squared fluctuation in units of mK$^2$:
\begin{equation}\label{eqn:PK}
    \Delta^2 = \frac{k^3}{2\pi^2}P_{21}(k) \langle T_{21}\rangle^2\ , 
\end{equation}
where $k^3P_{21}(k)/{2\pi^2}$ is the dimensionless squared fluctuation.

In order to calculate the large-scale 21-cm brightness temperature field, we combine the small-scale hydrodynamical simulations with a large-scale realization of the Universe in a large cubic volume using our 21-cm Semi-numerical Predictions Across Cosmic Epochs \citep[21cmSPACE,][]{visbal12, fialkov14, Reis2020} simulation code.  By interpolating (and slightly extrapolating) the tabulated 21-cm signal for the 17 combinations of density and streaming velocity, we calculate the 21-cm signal for every pixel of the large-scale grid (see Appendix \ref{sec: method_interp}). This approach allows us to account for both large-scale fluctuations (from three to up to hundreds of Mpc) as well as small-scale non-linear structure formation on sub-Mpc scales. We then evaluate the 21-cm power spectrum using a large-scale box that contains a (1536 Mpc)$^3$ comoving volume ($512^3$ grid). This allows us to accurately predict the power spectrum over the range $k = 0.01 - 1\ \rm{Mpc}^{-1}$. We use a $256^3$ grid for some minor comparison cases, and also go down to $128^3$ to test the effect of large-scale resolution. 

We note that we restore the 21-cm signal of each pixel by multiplying by its mean density (since the Lagrangian to Eulerian conversion for SPH, of division by density, is appropriate for small scales that are averaged over, but we do not want to erase the large-scale fluctuations). To clarify, our use of 21cmSPACE code in the present work is quite limited. We obtain the linear density and velocity cube from 21cmSPACE, but then derive the resulting 21-cm intensity cube at each redshift using the interpolation of the hydrodynamical simulations. Note that we match the linear $\delta$ of the total matter density to the initial conditions of the hydrodynamical simulations. Also, we account for the anisotropic redshift space distortion due to the large-scale velocity field (as detailed in the above references on the 21cmSPACE code), but in this paper we only plot the final spherically-averaged power spectrum. 

\section{Results}\label{sec:results}

\subsection{Main results}

We present the results for three different scenarios of Lyman-$\alpha$ coupling, characterized by the standard $x_{\alpha}$ parameter \citep{barkana05}, which quantifies the strength of the coupling. These scenarios are denoted as follows: 
\begin{itemize}
    \item \textbf{Dark Ages ($x_{\alpha}=0$):} the cosmological case with no astrophysical Lyman-$\alpha$ radiation;

    \item \textbf{Saturated coupling ($x_{\alpha}\rightarrow \infty$):} the limiting case where the Lyman-$\alpha$ radiation is intense enough to fully couple the 21-cm spin temperature to the kinetic gas temperature;

    \item \textbf{Moderate Coupling ($x_{\alpha}=1$):} an intermediate case where the coupling is significant but not close to saturated; commonly used to define the characteristic moment of the Lyman-$\alpha$ coupling transition. 
\end{itemize}
We note that in the case of Moderate Coupling, we would expect some contribution to the power spectrum from spatial fluctuations in the Lyman-$\alpha$ intensity \citep{barkana05}. We leave that for future work, and here we focus on the 21-cm fluctuations due to fluctuations in the gas density, temperature, and velocity. In the other cases we consider (Dark Ages, and Saturated coupling), there are no 21-cm fluctuations induced by Lyman-$\alpha$ fluctuations.

It is not known when the transition from the Dark Ages to Cosmic Dawn occurred in the real Universe, so we consider the various cases in the redshift range 20-40. In plots we include higher redshifts (up to 75) for the Dark Ages and lower redshifts (down to 15) for the Saturated coupling case.

To compare with a realistic observational sensitivity, we use the upcoming SKA, with specifications as shown in Table~\ref{tab:table1}. We assume 1080 hours of total integration time, optimistic foreground removal (required in order to reach the lowest wavenumbers), and consider the configurations AA$^\star$ (the current plan for SKA-Low, to be fully operational within $\sim 5$ years) and AA4 (a future upgrade). 

\begin{table}
\centering
\begin{tabular}{lc} 
\hline
Observational parameters   & Assumed values \\ 
\hline\hline
SKA configuration   & AA$^\star$, AA4  \\
Number of days      & 180.0 \\
Time per day (hrs)  & 6.0   \\
Foreground removal model    & Optimistic \\
Contribution        & Thermal variance \\

\hline
\end{tabular}
\caption{Observational configuration and parameters for the SKA. We calculated the sensitivity using the publicly-available code 21cmSense \citep{Pober2013, Pober2014}, assuming a frequency bandwidth of 8~MHz and $k$ bins of $\Delta \ln k = 0.296$.  We include thermal noise but not sample variance; for the redshifts and wavenumbers of interest here, cosmic variance is small and is not a significant barrier to detection.}
\label{tab:table1}
\end{table}

\begin{figure*}
    \centering
    \includegraphics[scale=0.5]{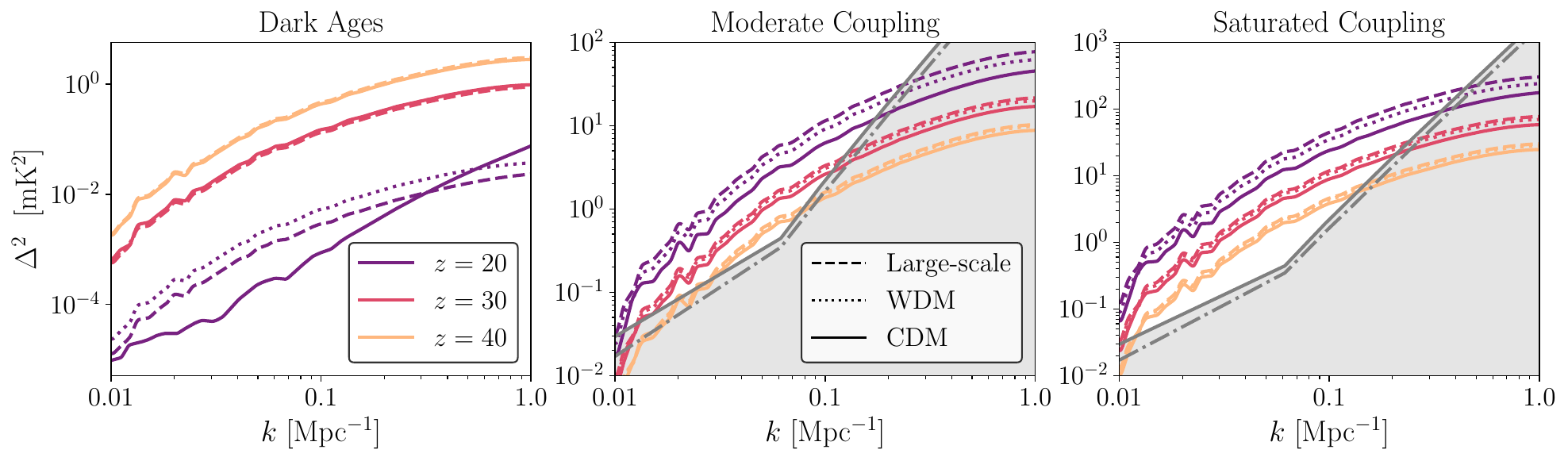}
    \includegraphics[scale=0.5]{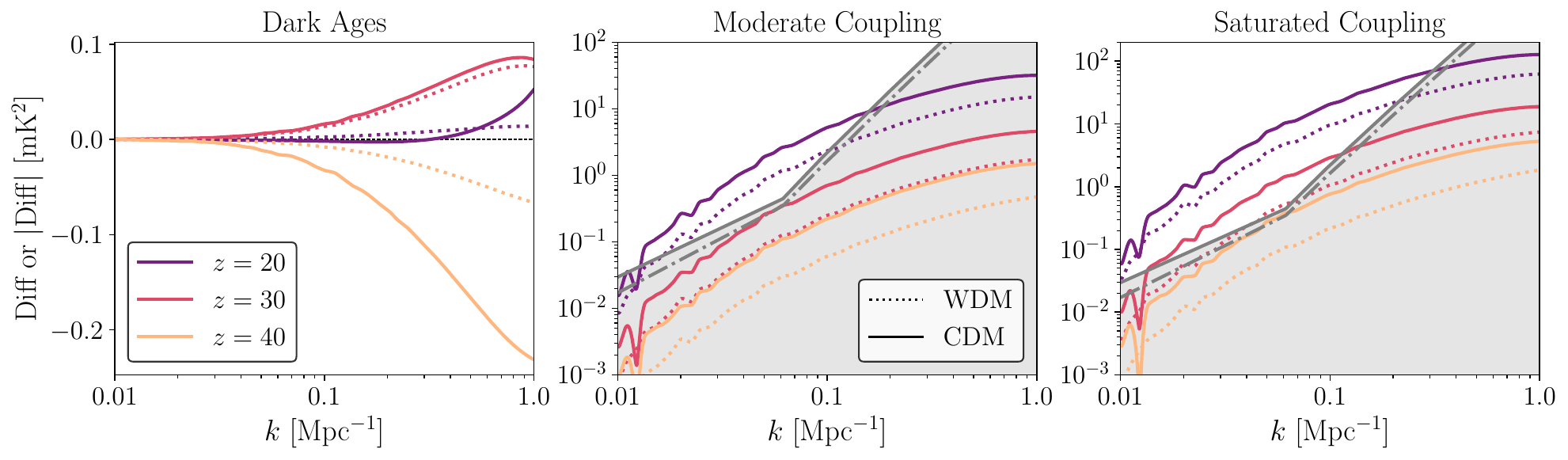}
    \caption{\textbf{Top panels:} The 21-cm power spectrum as a function of $k$, for the CDM model (solid), WDM-like model with $k_{\rm{cut}} = 100 \ h\ \rm{Mpc}^{-1}$  (dotted) and Large-scale fluctuations only (dashed). We consider the Dark Ages, as well as the Lyman-$\alpha$ coupling cases of Moderate coupling and Saturated coupling, each at $z = 20, 30$ or 40.   \textbf{Bottom panels:} The difference "Diff" in $\Delta^2$ due to clumping (i.e., between a given model and the case of Large-scale fluctuations only), as a function of $k$. Panels with coupling show $|\rm{Diff}|$; those panels also show the $z=20$ sensitivity for SKA AA$^\star$ (solid grey line and the corresponding shaded area) and SKA AA4 (dot-dashed grey line). We note that the precise output redshifts that we used from the numerical simulations are 19.46, 30.00 and 39.89 (which we loosely refer to as 20, 30, and 40).}\label{fig:21cmPS_vs_k_3cases}
    
\end{figure*}

In general, the 21-cm power spectrum is a powerful data set that is a function of $k$ and $z$. In plots, we show either the power spectrum versus $k$ at several redshifts, or versus $z$ at several wavenumbers. The top panels of Fig.~\ref{fig:21cmPS_vs_k_3cases} show $\Delta^2$ as a function of $k$ for three redshifts, $z=20, 30$ and 40. The 21-cm power spectrum is shown for three models: standard CDM, the WDM-like model, and the case of large-scale fluctuations only that excludes small-scale clumping. In the latter case, we include large-scale density fluctuations on our grid but use the results from uniform density small simulations boxes; note that in this case the streaming velocity has no effect. The case of large-scale fluctuations can also be seen approximately as the linear perturbation limit (since the fluctuations are fairly linear on the grid scale and above, at the plotted redshifts).

In the Dark Ages, the power spectrum has been previously calculated analytically \citep{2014alihamoud,2023NatAs...7.1025M}, in the linear regime (but see the discussion below in connection with Fig.~\ref{fig:PK_vs_kz_5cases}). The amplitude of the signal is relatively small, since collisional coupling (in the absence of Lyman-$\alpha$ photons) is weak at lower redshifts, while the density fluctuations are small at higher redshifts. The effects of clumping (as shown by the differences between the models) are small but significant. A few illustrative values of the power spectrum are shown in Table~\ref{tab:table2}. For example, at $z=30$ and $k=0.05$~Mpc$^{-1}$, clustering enhances the 21-cm power spectrum by 13\% in the case of CDM (i.e., CDM compared to large-scale fluctuations only), with the enhancement reduced to 10\% for WDM-like. 

\begin{table*}
\centering
\begin{tabular}{cclll} 
\hline
21-cm observable  & Model & Dark Ages ($z=30$) & Moderate Coupling ($z=20$) & Saturated Coupling ($z=20$)\\ 
\hline\hline

Power spectrum  & Large-scale  & $4.78\times 10^{-2}$ & 4.22  \verb|{|13.0\verb|}| & 16.4  \verb|{|50.5\verb|}| \\
\verb|[|mK$^2$\verb|]|          & WDM   & $5.28\times 10^{-2}$ \verb|[|+10.5\%\verb|]|& 3.36 \verb|[|-20.4\%\verb|]| \verb|{|10.3\verb|}|& 12.9 \verb|[|-21.5\%\verb|]| \verb|{|39.7\verb|}|\\
                & CDM   & $5.40\times 10^{-2}$ \verb|[|+13.0\%\verb|]|& 2.32 \verb|[|-45.0\%\verb|]| \verb|{|7.14\verb|}| & 8.93 \verb|[|-45.6\%\verb|]| \verb|{|27.5\verb|}|\\
                & SKA AA$^\star$ & & 0.325 & 0.325 \\
                & SKA AA4 & & 0.246 & 0.246 \\
                
\hline
Power spectrum & Large-scale  & $2.64\times 10^{-3}$ & $4.54\times 10^{-4}$ & $4.23\times 10^{-4}$ \\
\verb|[|dimensionless\verb|]|         & WDM   & $2.59\times 10^{-3}$ \verb|[|-1.89\%\verb|]| & $3.88\times 10^{-4}$ \verb|[|-14.5\%\verb|]| & $3.56\times 10^{-4}$ \verb|[|-15.8\%\verb|]| \\
                & CDM   & $2.44\times 10^{-3}$ \verb|[|-7.58\%\verb|]| & $2.93\times 10^{-4}$ \verb|[|-35.5\%\verb|]| & $2.71\times 10^{-4}$ \verb|[|-35.9\%\verb|]| \\

\hline
\end{tabular}

\caption{A few illustrative values of the 21-cm power spectrum, as observed (in mK$^2$) and dimensionless (i.e., the power spectrum divided by the squared global signal), at $k = 0.05$ Mpc$^{-1}$, for the Dark Ages, Moderate Coupling and Saturated Coupling cases. For the WDM-like and CDM models we also indicated in square brackets the relative change in each case compared to the corresponding Large-scale case (i.e., the effect of clumping expressed as a percentage change). We also indicate (where relevant) the sensitivity of the SKA (AA$^\star$ and AA4 configurations), and in curly brackets the ratio of the signal to the SKA AA$^\star$ sensitivity. We note that the precise output redshifts that we used from the numerical simulations are 19.46 and 30.00 (which we loosely refer to as 20 and 30).}
\label{tab:table2}
\end{table*}

In order to gain physical understanding, it is useful to separate out the clumping effect on the relative size of fluctuations from the effect on the global signal (which plays a role in the 21-cm power spectrum as it is observed, in units of mK$^2$). Thus, the table also lists values of the dimensionless power spectrum. By this measure, clumping now {\it reduces}\/ the squared fluctuation, by 7.6\% in CDM. In general, the Dark Ages potentially present a clean signal, free from astrophysics, but the weak signal and the need to go to space to avoid the ionosphere (at $z \gtrsim 30$) imply that detection requires futuristic lunar-based telescope arrays with very large collecting areas \citep{2023NatAs...7.1025M}.

Once Lyman-$\alpha$ coupling kicks in due to the first stars, the amplitude of the power spectrum increases compared to the Dark Ages on all scales. Saturated coupling gives a higher power spectrum amplitude than moderate coupling, as the Lyman-$\alpha$ coupling makes the 21-cm signal fully sensitive to the gas temperature (which is colder than the CMB on average). Again considering $k=0.05$~Mpc$^{-1}$ (but now at $z=20$), with Moderate Coupling, clumping lowers the power spectrum by 45\% in CDM and 20\% in $100h$ WDM-like; for the dimensionless power spectrum (i.e., the effect on fluctuations without including the effect on the global signal), the reductions are somewhat smaller. With Saturated Coupling, clumping has a similar relative effect (see Table~\ref{tab:table2} for the precise numbers). Thus, in both cases, {\it clumping reduces the observable power spectrum by (nearly) a factor of two}.\/ The WDM-like cutoff model has a clumping impact that is less than half of the effect in CDM; thus, clumping can be used to constrain the dark matter model or particle mass if there is a cutoff in this range of scales.

In order to highlight the effect of clumping, in the bottom panels of Fig.~\ref{fig:21cmPS_vs_k_3cases} we show the difference between the 21-cm power spectrum in the CDM or WDM-like model and that in the large-scale (no clumping) case. In the Dark Ages, the effect of clumping can have either sign, while in the cases with coupling, the effect is always negative, so in those cases we show the absolute value of the difference. If these levels of coupling are reached by $z=20$, then the effect of clumping can potentially be observed with the SKA. 
We illustrate at $k = 0.05$ Mpc$^{-1}$ (which is close to the best $k$ for detection). The CDM power spectrum is higher than the SKA AA$^\star$ sensitivity by a factor of 27 for Saturated Coupling (see additional signal-to-noise ratios in Table~\ref{tab:table2}). More interesting is the ratio of the clumping effect to the SKA sensitivity. In CDM this is 5.8 (Moderate Coupling) or 23 (Saturated Coupling). For detecting the difference between WDM-like and CDM (with clumping included), the ratio is 3.2 (Moderate Coupling) or 12 (Saturated Coupling). SKA AA4 would improve all of the signal-to-noise ratios by a factor of 1.3. Note that in the coupled cases, WDM-like has a smaller clumping effect, which actually makes this model easier to detect than CDM (since clumping suppresses the 21-cm power spectrum in these scenarios).

\begin{figure*}
    \centering
    \includegraphics[scale=0.5]{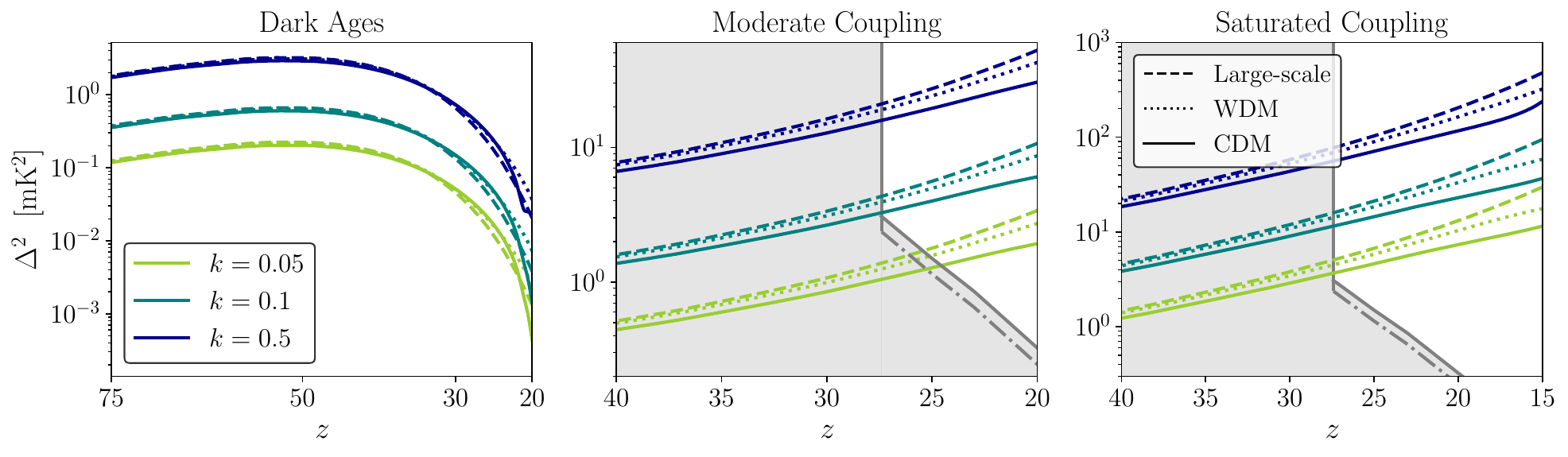}
    \includegraphics[scale=0.5]{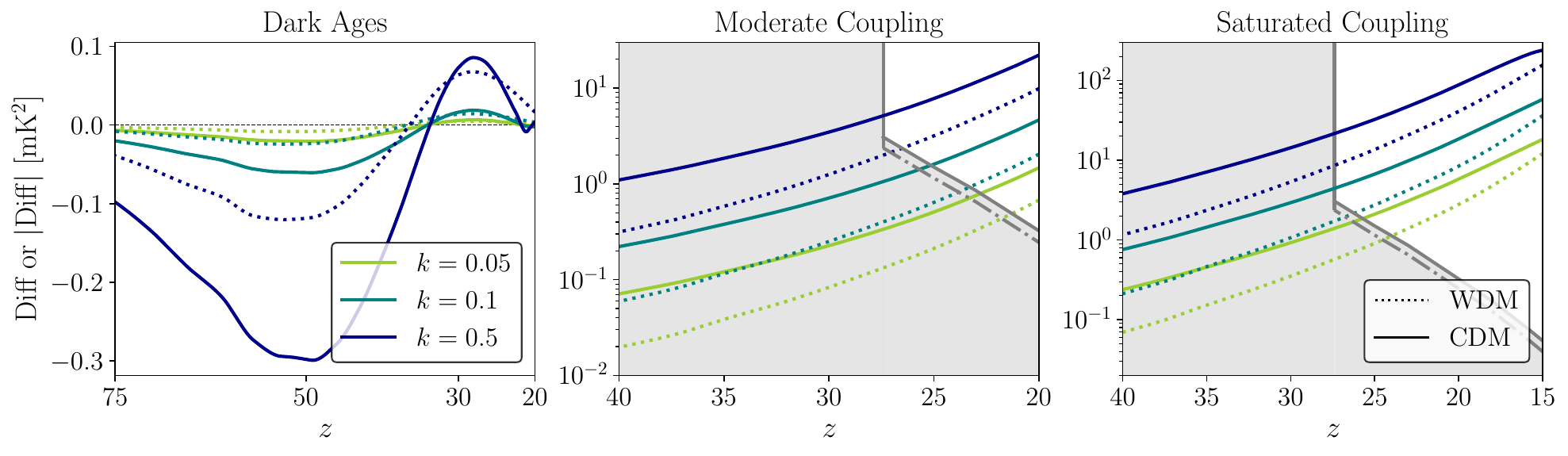}
    \caption{\textbf{Top panels:} The 21-cm power spectrum as a function of $z$ at $k = 0.05$, 0.1 and 0.5 Mpc$^{-1}$. We show the same models and cases as in Fig.~\ref{fig:21cmPS_vs_k_3cases}: CDM (solid), WDM-like (dotted), and Large-scale (dashed), for Dark ages, Moderate coupling and Saturated coupling. \textbf{Bottom panels:} The difference "Diff" in $\Delta^2$ due to clumping. Panels with coupling show $|-\rm{Diff}|$, and also show the $k= 0.05$ Mpc$^{-1}$ sensitivity for SKA AA$^\star$ (solid grey line and the corresponding shaded area) and SKA AA4 (dot-dashed grey line).}\label{fig:21cmPS_vs_z_3cases}
\end{figure*}

We show the other cut of the 21-cm power spectrum in Fig.~\ref{fig:21cmPS_vs_z_3cases}. The top panels display the 21-cm power spectrum as a function of $z$ for three values of $k$: 0.05, 0.1, and 0.5 Mpc$^{-1}$. We again show the power spectrum for CDM, $100h$ WDM-like, and the large-scale (no clumping) case. The bottom panels again show the effect of clumping, in terms of the power spectrum in each model relative to the large-scale no-clumping case. 

The effect of non-linear clumping evolves with cosmic time. Starting with the Dark Ages and standard CDM, the power spectrum behaves similarly at the various wavenumbers shown; we illustrate here the values at $k=0.5 \ \rm{Mpc}^{-1}$. The power spectrum is suppressed at $z = 75$ by 5.7\%. Below $z=75$, the difference in the power spectrum becomes more negative, reaching a maximum negative $\rm{Diff}=-0.30$ mK$^2$ (a 10.4\% suppression) at $z=49$. After $z=49$, the difference becomes less negative, and after $z=32$, the difference turns positive, meaning that clumping now increases the power spectrum compared to the large-scale no-clumping case. The positive difference in the power spectrum reaches a maximum $\rm{Diff}=0.086$ mK$^{2}$ (a 17\% enhancement) at $z=28$, after which it declines towards zero as collisional coupling weakens globally, reducing all aspects of the 21-cm signal. 

At smaller wavenumbers, the fluctuations are smaller and so is the (absolute) effect of clumping. For CDM at $z \sim 49$, the maximum difference due to clumping is again 10\% at both $k=0.1$ and 0.05 Mpc$^{-1}$. The maximum enhancement at $z \sim 28$ is 19\% at both of these wavenumbers. In the WDM-like model, where clumping is less pronounced due to weaker structure formation, the maximum enhancement at $z \sim 28$ is $14-15\%$ at $k = 0.05-0.5$~Mpc$^{-1}$. An important point is that the clumping significantly affects the power on a wide range of length scales, and the fractional effect varies slowly with $k$ over the range of scales that we probe.


In the early Cosmic Dawn cases, when stellar radiation provides substantial Lyman-$\alpha$ coupling, clumping consistently suppresses the 21-cm power spectrum at all redshifts, for both the WDM-like and CDM models. The absolute clumping effect in the power spectrum increases with time and with wavenumber (along with similar behavior by the power spectrum itself). For Moderate Coupling, at $z=20$ and $k=0.5 \ \rm{Mpc}^{-1}$, the 21-cm power spectrum is suppressed by 43\% in CDM, and 19\% in the WDM-like model. For Saturated Coupling, these numbers change to $49\%$ for CDM and 33\% for the WDM-like model. These numbers are fairly close to the values listed earlier for the percentage suppression at $k=0.05$~Mpc$^{-1}$. The relative suppression decreases with redshift, so that (e.g., for Saturated Coupling at $k=0.05$~Mpc$^{-1}$ in CDM) it goes from 46\% at $z=20$ to 24\% at $z=30$ and 16\% at $z=40$, while it is as high as 61\% at $z=15$.

\subsection{Physical explanation}

We can qualitatively understand the various contributions to the effect of clumping in the various cases. This is surprisingly complex and subtle. We wish to understand the effect of small-scale density fluctuations (which we have called ``clumping") on large-scale 21-cm fluctuations. The first step is to consider a single large-scale region. 

Suppose such a region has some given uniform density, which yields a certain 21-cm intensity. Now consider adding clumping, i.e., adding internal small-scale density fluctuations while keeping the overall mean density the same (in keeping with mass conservation). Analyzing the effect of such clumping on the resulting overall 21-cm intensity is qualitatively similar to the effect of clumping on the global 21-cm signal, as long as we consider a large-scale region that is much larger than the scale of the clumping. While clumping maps some gas into hot, high-density clumps, it leaves other gas within cold, low-density voids; the overall effect is not at all obvious. Now, linear density fluctuations cannot change the overall 21-cm signal, since they are symmetric between positive and negative density fluctuations. If we include only the effect of density (and not temperature or 21-cm coupling), this holds true even for non-linear density fluctuations, since dense clumps take on a smaller volume, and the volume-averaged 21-cm intensity is unchanged, since the mean density of the region is unchanged. We note though that the 21-cm signal is not an exact volume average, due to the complicating effect of redshift distortions along the line of sight (which slightly enhance the 21-cm effect of density). Still, the main effect on the mean 21-cm signal comes from the combination of non-linear density fluctuations with temperature and coupling fluctuations, and the resulting asymmetry between positive and negative fluctuations. Positive, high-density fluctuations are more strongly non-linear, with gas reaching high overdensities and heating up more substantially (including in shocks) than underdense gas is cooled. When there is substantial outside (Lyman-$\alpha$) coupling, this asymmetry dominates the overall mean, and reduces the overall absorption signal. It overcomes effects favoring low temperatures, including the stronger sensitivity of $T_{21}$ to low gas temperatures due to the asymmetric dependence $\propto (T_S - T_{\rm CMB})/T_S$; and the reduced effect of high temperatures due to saturated heating (when $T_S \gg T_{\rm CMB}$, which is rare even for highly overdense gas at the redshifts relevant to this work). 

The Dark Ages case is even more complex due to the strong positive correlation of coupling (which is collisional in that case) with density, which increases the intensity of the 21-cm signal (which itself is a negative, absorption signal) and can reverse the sign of the overall effect of clumping. We found in \citet{Park} a suppression of the global signal (i.e., a reduction in the 21-cm absorption) at all redshifts with moderate or saturated coupling, while the Dark Ages case had a suppression at high redshift and an enhancement at redshifts below $z=46$, with a maximum enhancement at $z=27$ (where these redshift values are for CDM). More details can be found in \citet{Park}, including the distribution of gas densities and spin temperatures (see the Supplementary note for detailed figures), and the small additional modification in the case of the global 21-cm signal due to the averaging over large-scale regions.

The qualitative discussion thus far focused on a single large region. To understand the effect on large-scale fluctuations, we need to compare different large-scale regions and consider their 21-cm brightness temperature difference (which is the fluctuation). We can do this in some detail in 3~Mpc regions (the size of our hydrodynamical simulation box size, or a single pixel in our large-scale grid). Consider $z=19.5$ (an output redshift of the hydro simulations) and Saturated Coupling. Compare the mean density box $0\, \sigma_\star$, with typical plus and minus density fluctuation regions, $1\, \sigma_\star$ and $-1\, \sigma_\star$ (see section~\ref{s:sim}). Comparing such uniform regions (i.e., without clumping), the density fluctuation is $\pm 13.8\%$ in linear theory, and +16.0\% and $-12.9\%$ non-linearly (in terms of the actual comparison of the $1\, \sigma_\star$ and $-1\, \sigma_\star$ boxes, respectively, to $0\, \sigma_\star$). Since the effect of gas temperature goes in the opposite direction of density (i.e., increased density strengthens the absorption, while the hotter gas reduces it), the 21-cm fluctuation is only +6.5\% and $-4.5\%$ in magnitude (e.g., the $1\, \sigma_\star$ box has stronger 21-cm absorption by 6.5\% than the mean density box). When we add clumping to all the boxes, the mean density 21-cm intensity is reduced in magnitude by 8.7\%. The $1\, \sigma_\star$ box is reduced by 11.5\%, and the $-1\, \sigma_\star$ box is reduced by 7.0\%. This shrinks the relative 21-cm fluctuation to +3.2\% and $-2.7\%$. In other words, since the overall large-scale density is positively correlated with absorption magnitude, and clumping has an effect (of reducing the magnitude) that increases with large-scale density, its overall impact is to reduce the 21-cm fluctuation. We can also intuitively understand why the effect of clumping increases with large-scale density: a region of higher mean density is equivalent to a higher-density universe, where cosmic expansion is slower, the dynamical time is shorter, and processes of structure formation progress more rapidly.

We note that, when comparing the clumping effect on the power spectrum and on the global signal, the former is larger, but the difference is not as substantial as it may at first appear. For example, consider the case from Table~\ref{tab:table2} that shows a 46\% reduction in the power spectrum ($z=20$, $k=0.05$~Mpc$^{-1}$, Saturated Coupling, CDM). If we examine the dimensionless power spectrum (i.e., after the global signal was factored out), the reduction is only 36\%. Since the power spectrum represents squared fluctuations, the reduction in the size of the relative fluctuation is 17\%. In comparison, the effect of clumping on the global signal (in the same case) is a reduction (in magnitude) of 7.9\%.

\subsection{Streaming velocity and large-scale convergence}

\begin{figure*}
    \centering
    \includegraphics[scale=0.5]{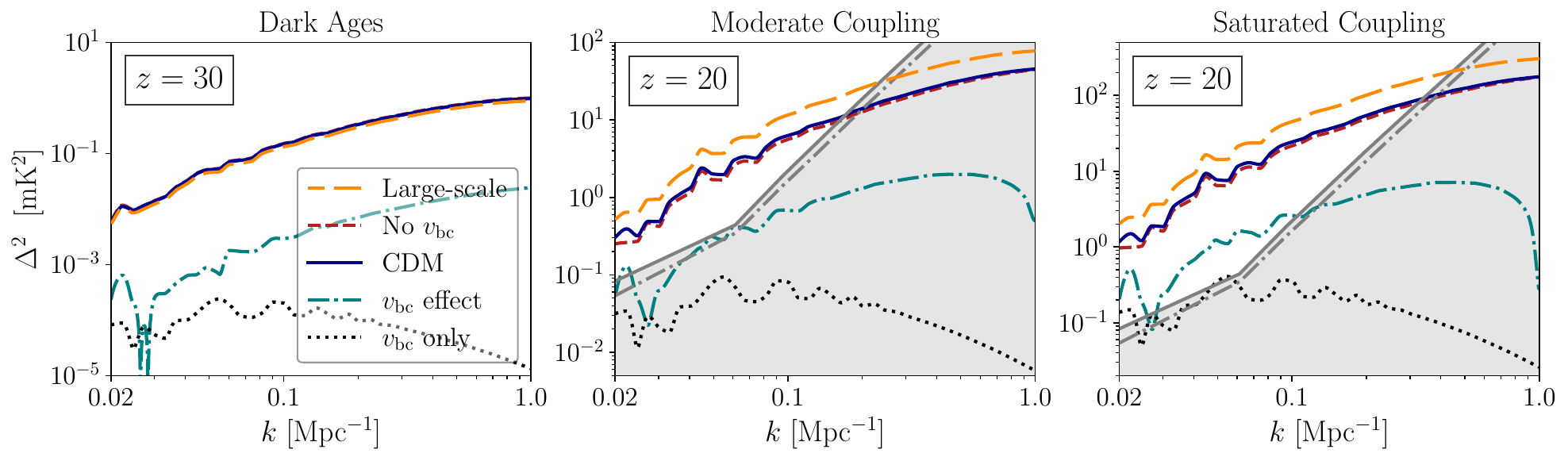}
    \includegraphics[scale=0.5]{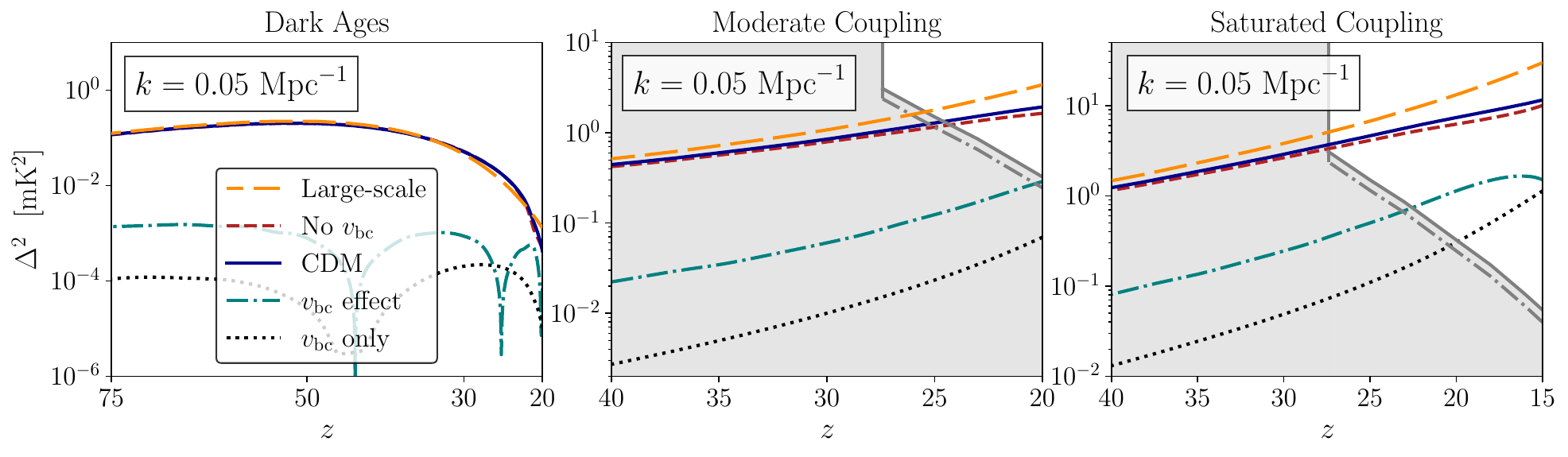}
    \caption{\textbf{Top panels:} 21-cm power spectrum as a function of $k$ for the CDM (solid) and Large-scale fluctuation (long-dashed) cases (repeated from the top panels of Fig.~\ref{fig:21cmPS_vs_k_3cases}, but shown here only down to $k= 0.02$ Mpc$^{-1}$), along with the No~$v_{\rm{bc}}$ (short-dashed), $v_{\rm{bc}}$~effect (dot - dashed, shown in absolute value), and $v_{\rm{bc}}$~only (dotted) cases (see text). Results are shown for the Dark Ages at $z = 30$, and for Moderate coupling and Saturated coupling at $z = 20$. The panels with coupling show the $z= 20$ sensitivity for SKA AA$^\star$ (solid grey line and corresponding shaded area) and SKA AA4 (dot-dashed grey line). We note that the precise output redshifts that we used from the numerical simulations are 19.46 and 30.00 (which we loosely refer to as 20 and 30). The $v_{\rm{bc}}$~effect is positive in the coupling cases, and for the Dark Ages case it is positive in the $k$ range [0.026, 0.028] and negative otherwise. \textbf{Bottom panels:} 21-cm power spectrum as a function of $z$ for the CDM and Large-scale fluctuation cases (repeated from the top panels of Fig.~\ref{fig:21cmPS_vs_z_3cases}), as well as the No~$v_{\rm{bc}}$, $v_{\rm{bc}}$~effect,  and $v_{\rm{bc}}$~only cases. We consider the Dark ages, Moderate coupling, and Saturated coupling, all at $k = 0.05$ Mpc$^{-1}$. The panels with coupling show the $k= 0.05$ Mpc$^{-1}$ sensitivity for SKA AA$^\star$ and SKA AA4. The styles of the cases and sensitivities are the same as in the top panels. The $v_{\rm{bc}}$~effect is positive in the coupling cases, and for the Dark Ages case it is positive in the $z$ ranges [20.17, 25.19] and [43.88, 75], and negative otherwise. Note: In this figure, all cases were calculated with a $256^3$ large-scale grid, unlike the previous two figures, which used $512^3$.} \label{fig:PK_vs_kz_5cases}
    
\end{figure*}

In order to understand the role of the streaming velocity, and compare to previous work on that issue, in Fig.~\ref{fig:PK_vs_kz_5cases} we show the 21-cm power spectrum for two previously-shown cases (CDM and Large-scale fluctuations) alongside two additional cases in which some contributions are artificially turned off: the `No $v_{\rm{bc}}$' and `$v_{\rm{bc}}$ only' cases (where $v_{\rm{bc}}$ denotes the relative baryon --- CDM velocity). The `No $v_{\rm{bc}}$' case uses the results of small simulation boxes in which the mean density is varied but we set $v_{\rm{bc}}=0$; thus, it includes large-scale density fluctuations but excludes the streaming velocity. The `$v_{\rm{bc}}$ only' case is roughly the opposite, with the density of each small simulation box set to the cosmic mean density but with a varying streaming velocity. Both of these cases include small-scale non-linear clumping, and are in the CDM model. These cases allow us to compare two different estimates of the effect of the streaming velocity. The `$v_{\rm{bc}}$ only' case estimates the separate effect of streaming velocities independently of large-scale density fluctuations; we refer to this as the naive effect of streaming. On the other hand, the difference between the full CDM case and the `No $v_{\rm{bc}}$' case yields the actual effect of streaming velocities, including the full non-linear interactions among the large-scale density and velocity fields and the small-scale clumping; we refer to this as the full effect of streaming, and also add it to the plot as the $v_{\rm{bc}}$~effect (defined as CDM minus No~$v_{\rm{bc}}$).

In the top panels of Fig.~\ref{fig:PK_vs_kz_5cases}, we show the power spectrum as a function of $k$ at $z = 30$ for the Dark Ages, and at $z = 20$ for the Moderate Coupling and Saturated Coupling scenarios (the overall setup is similar to that of the top panels of Fig.~\ref{fig:21cmPS_vs_k_3cases}). The bottom panels of Fig.~\ref{fig:PK_vs_kz_5cases} show the 21-cm power spectrum as a function of $z$ for the same models/cases, at $k = 0.05 \ \rm{Mpc}^{-1}$ (the overall setup is similar to that of the top panels of Fig.~\ref{fig:21cmPS_vs_z_3cases}). 

The naive effect of streaming in the Dark Ages was estimated by \citet{2014alihamoud}, who analytically calculated the modulation by $v_{\rm{bc}}$ of the linear power spectrum, then found the contribution of this modulation to the large-scale 21-cm fluctuations to second order. Compared to our fully non-linear simulations, their calculation did not include non-linear effects that occur in the absence of $v_{\rm{bc}}$ (which we find are the dominant contribution of clumping), and their $v_{\rm{bc}}$ calculation used a combination of linear and second-order approximations. We find a similar order of magnitude of the naive effect. For example, at $z=30$ and $k = 0.05 \ \rm{Mpc}^{-1}$, we find a naive velocity effect (from Fig.~\ref{fig:PK_vs_kz_5cases}) that is 0.46\% of the large-scale (i.e., no clumping) signal; this can be compared to about 2\% in Fig.~14 of \citet{2014alihamoud}. Some of this discrepancy is alleviated by noting that in Fig.~14 of \citet{2014alihamoud} the plotted monopole power spectrum does not include the effect of redshift-space distortions (line-of-sight velocity gradients) which increase the power spectrum by nearly a factor of two \citep{SB_SSA_2004,barkana2005}.

To understand the importance of the various effects, we compare at $k = 0.05 \ \rm{Mpc}^{-1}$ the naive velocity effect, the full effect of streaming velocity, and the total effect of clumping, all as a percentage of the large-scale signal. For the Dark Ages at $z=30$, the naive velocity effect is 0.46\%, the full velocity effect is -1.9\%, and the total clumping effect is 13\%. At $z=20$, these numbers for Moderate (or Saturated) coupling are 2.2\% (2.5\%), 8.4\% (8.5\%), and 46\% (46\%), respectively. More generally, from Fig.~\ref{fig:PK_vs_kz_5cases} we can see that the full velocity effect is larger than the naive velocity effect by between 0.5 and 2 orders of magnitude, depending on $z$ and $k$, yet is still about an order of magnitude smaller than the overall effect of clumping. Note that the numbers listed here from Fig.~\ref{fig:PK_vs_kz_5cases} were calculated with a $256^3$ large-scale grid, unlike the previous two figures, which used $512^3$.

\begin{figure*}
    \centering
    \includegraphics[scale=0.5]{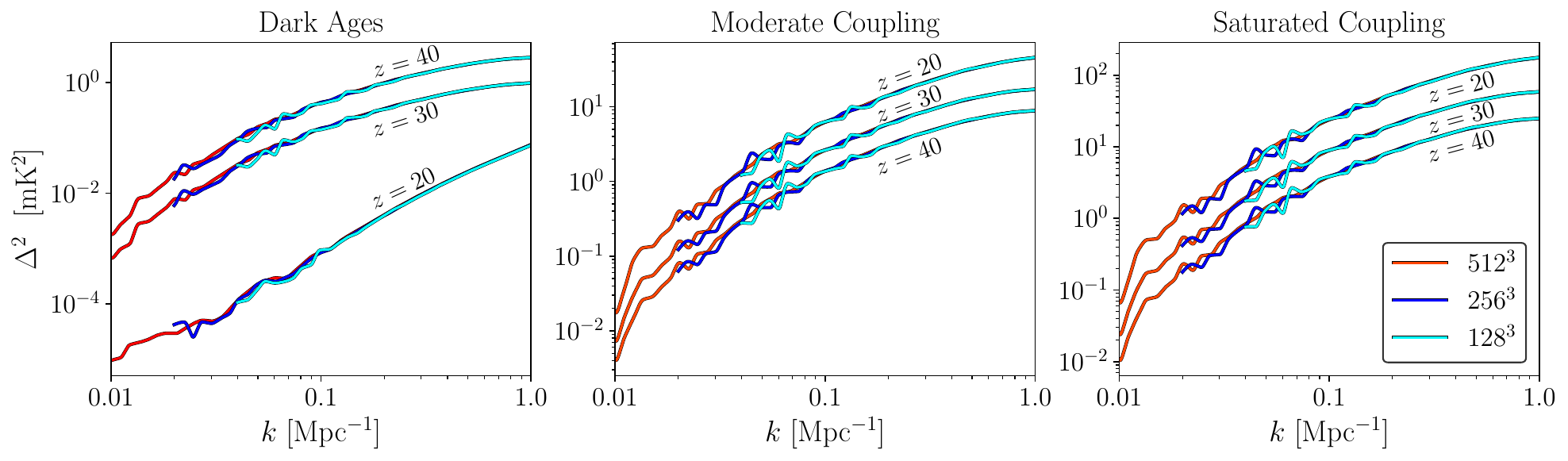}
    \caption{Testing convergence with respect to the size of our large-scale grid. We show the 21-cm power spectrum of CDM as a function of $k$ at $z=20$, 30 and 40, for three box sizes: 1536, 768 and 384 Mpc on a side, corresponding to $512^3, 256^3$ and $128^3$ pixels, respectively. The curves go down to  $k = 0.01$, 0.02, and 0.04 Mpc$^{-1}$, respectively. We note that the precise output redshifts that we used from the numerical simulations are 19.46, 30.00 and 39.89 (which we loosely refer to as 20, 30, and 40).}\label{fig:power_spectrum_box_size}
\end{figure*}

Finally, we consider scale convergence in some detail. We have previously \citep{Park} shown the approximate convergence of our results with respect to small-scale resolution of the hydrodynamic simulations. Here we test convergence with respect to the size of the large-scale grid. In Fig.~\ref{fig:power_spectrum_box_size}, we present the 21-cm power spectrum as a function of $k$, computed for simulation box sizes of 384 Mpc (teal curves), 768 Mpc (blue curves), and 1536 Mpc (red curves) at redshifts $z=20, 30$ and 40. These correspond to $128^3$, 256$^3$, and $512^3$ pixels (voxels), with a fixed side length of 3~Mpc. In our previous results, we used our largest, $512^3$ grid (except for the `No $v_{\rm{bc}}$' and `$v_{\rm{bc}}$ only' cases in Fig.~\ref{fig:PK_vs_kz_5cases}, where we used 256$^3$). We plot the curves down to $k= 0.01$~Mpc$^{-1}$ for $512^3$, 0.02~Mpc$^{-1}$ for 256$^3$, and 0.04~Mpc$^{-1}$ for $128^3$.

These power spectra, calculated for the CDM model, exhibit consistent results across the overlapping ranges of wavenumbers $k$. This consistency indicates that the cosmological processes driving the 21-cm signal are well captured across all box sizes.
Increasing the box size enables probing of larger physical scales (smaller $k$ values), and reduces sample variance. For the Dark Ages case, for example, the power spectra for $128^3$, at the three redshifts shown, are accurate to within 5\% (compared with the result for 256$^3$) at $k> 0.15$~Mpc$^{-1}$, and to within 25\% at $k> 0.06$~Mpc$^{-1}$. The 256$^3$ result (compared to $512^3$) is accurate to within 5\% at $k> 0.09$~Mpc$^{-1}$, and to within 20\% at $k> 0.03$~Mpc$^{-1}$. Extrapolating, we expect that our final result with $512^3$ should be accurate to 5\% roughly at $k> 0.04$~Mpc$^{-1}$, and to 20\% roughly at $k> 0.015$~Mpc$^{-1}$. 

\section{Summary and conclusions}\label{sec:Conclusion}

We used a novel method that combines high-resolution small-scale hydrodynamical simulations with a large-scale semi-numerical grid, to explore the impact of non-linear small-scale structure formation (clumping) on the large-scale 21-cm power spectrum. With this method, we covered the range of scales from sub-Mpc up to Gpc. At one end, the WDM-like model shows that the median scale for the clumping effect is a radius (half-wavelength) of $\sim 60$~kpc, corresponding to a halo mass scale (i.e., the total mass contained within such an initial comoving radius) of $3\times 10^7 M_{\odot}$. At the other end, we used a grid of 1.536~Gpc on a side to predict the 21-cm power spectrum over the wavenumber range of $0.01-1$~Mpc$^{-1}$. Our convergence study suggests that our largest grid captures the power spectrum with a sample variance error that is below 5\% at $k> 0.04$~Mpc$^{-1}$, and below 20\% at $k> 0.015$~Mpc$^{-1}$. 

We considered the Dark Ages and early Cosmic Dawn ($z=20 - 40$), and included contributions to the power spectrum from fluctuations in the gas density, temperature, and velocity. We left for future work the contribution from Lyman-$\alpha$ fluctuations (which would be expected in our Moderate Coupling case). As the clumping effect arises from small-scale density fluctuations, it offers a unique opportunity to probe the standard CDM model in a new regime and thus potentially investigate the properties of dark matter.  To this end, we studied standard cold dark matter as well as a warm dark matter -- like model with a Gaussian cutoff above $k_{\rm{cut}} = 100 \, h \,\rm{Mpc}^{-1}$. We note that we study the direct effect of clumping, which arises from the non-linear, filamentary cosmic web; there is a separate indirect effect from Lyman-$\alpha$ coupling, which depends on star formation in the first galaxies, and has been suggested as a way to constrain small-scale matter clumping \citep{2020PhRvD.101f3526M}, though this is limited by large uncertainties in the astrophysics of star formation and stellar feedback.

We found that clumping has a significant impact on the 21-cm power spectrum, requiring a substantial correction to standard theoretical predictions. For example, for the Dark Ages case at $z=30$ and $k=0.05$~Mpc$^{-1}$, clustering enhances the 21-cm power spectrum by 13\% (corresponding to 0.0062~mK$^{2}$) in the case of CDM, with the enhancement reduced to 10\% for WDM-like. The maximum enhancement in CDM occurs at $z \sim 28$ and is 0.0066~mK$^{2}$ (a 19\% effect) at $k=0.05$~Mpc$^{-1}$. These effects are significant, but the signal itself is quite small in this case and its detection requires futuristic telescope arrays.

Most importantly, once Lyman-$\alpha$ coupling kicks in due to the first stars, the amplitude of the power spectrum increases, and also the effect of clumping becomes remarkably large. Considering $k=0.05$~Mpc$^{-1}$ at $z=20$, with Moderate Coupling, clumping lowers the power spectrum by 45\% (1.9~mK$^{2}$) in CDM and 20\% (0.86~mK$^{2}$) in WDM-like. With Saturated Coupling, clumping lowers the power spectrum by 46\% (7.5~mK$^{2}$) in CDM and 21\% (3.5~mK$^{2}$) in WDM-like. Thus, in both cases, clumping reduces the observable power spectrum by almost a factor of two, while the WDM-like cutoff can be distinguished from CDM since its clumping impact is less than half of the effect in CDM.

Although there are various physical contributions, we can
qualitatively understand these reductions due to clumping in cases
with strong outside coupling (i.e., coupling that is independent of
local density, supplied here by Lyman-$\alpha$ radiation). In any
given region, clumping {\it reduces}\/ the overall 21-cm absorption, due to
the asymmetry in which positive fluctuations are more strongly
non-linear, with gas reaching high overdensities and heating up more
substantially than underdense gas is cooled. Then, when comparing
regions with different large-scale densities, the size of the effect of clumping
increases with density, and this reduces the overall 21-cm
fluctuation.

These effects can in principle be observed with the SKA. Continuing with $k=0.05$~Mpc$^{-1}$ at $z=20$, the CDM power spectrum (including clumping) is higher than the sensitivity of the planned SKA AA$^\star$ configuration by a factor of 7.1 (Moderate Coupling) or 27 (Saturated Coupling); the clumping effect of CDM (i.e., the difference compared to large-scale fluctuations only) is higher than the SKA sensitivity by a factor of 5.8 (Moderate Coupling) or 23 (Saturated Coupling). The difference (which is entirely due to clumping) between WDM and CDM is 3.2 (Moderate Coupling) or 12 (Saturated Coupling) times the SKA sensitivity. 

In agreement with our study of the effect on the global signal \citep{Park}, we find that half the effect of clumping comes from scales below a radius (half-wavelength) of $\sim 56$~kpc, corresponding to a halo mass scale (i.e., the total mass contained within such an initial comoving radius) of $2.9\times 10^7 M_{\odot}$. This corresponds to a 6.6~keV WDM mass and a $2.2\times 10^{-20}$~eV FDM mass. Based on our previous work \citep{Park}, we expect that the range of scales probed by clumping (roughly the first and third quadrants of the effect) is $k_{\rm{cut}} \sim 30\, h - 300 \, h \, \rm{Mpc}^{-1}$,
which corresponds to radii\footnote{The various values listed in this paragraph are accurate; some values given in \citet{Park} were slightly off.} of $19 - 190$~kpc, and a halo mass scale of $1.1\times 10^6 - 1.1\times 10^9 M_{\odot}$. Note that although we quote equivalent halo mass scales, the clumping effect actually comes from the non-linear, filamentary cosmic web, and not primarily from virialized objects. These ranges imply that the clumping effect on the 21-cm signal can be used to constrain WDM masses in the range of $2.3-17$~keV and FDM masses in the range of $1.5\times 10^{-21} - 2.7\times 10^{-19}$~eV. 

In order to detect the clumping effect unambiguously during Cosmic Dawn, it will need to be distinguished from foregrounds as well as from astrophysical contributions to the 21-cm power spectrum. In particular, in the intermediate regime (between the Dark Ages and Saturated Coupling), spatial fluctuations in galaxy formation are also important, and they can be affected by small-scale fluctuations 
\citep{2020PhRvD.101f3526M} as well as astrophysical feedback.

\begin{acknowledgments}
SS and RB acknowledge the support of the Israel Science Foundation (grant no.\ 1078/24). HP was supported in part by grant NSF PHY-2309135 to the Kavli Institute for Theoretical Physics (KITP). Numerical simulations for this work were performed on the idark computing cluster of the Kavli Institute for Physics and Mathematics of the Universe, the University of Tokyo. NY acknowledges financial support from JSPS International Leading Research 23K20035. RB and NY acknowledge the JSPS Invitational Fellowship S24099.

\end{acknowledgments}




%
\software{\texttt{Numpy} \citep{harris2020array}, \texttt{Scipy} \citep{2020SciPy-NMeth}, \texttt{matplotlib} \citep{Hunter:2007}}



\appendix

\section{Interpolating the 21-cm brightness temperature}\label{sec: method_interp}

\begin{figure*}
\centering
\includegraphics[width=0.37\textwidth]{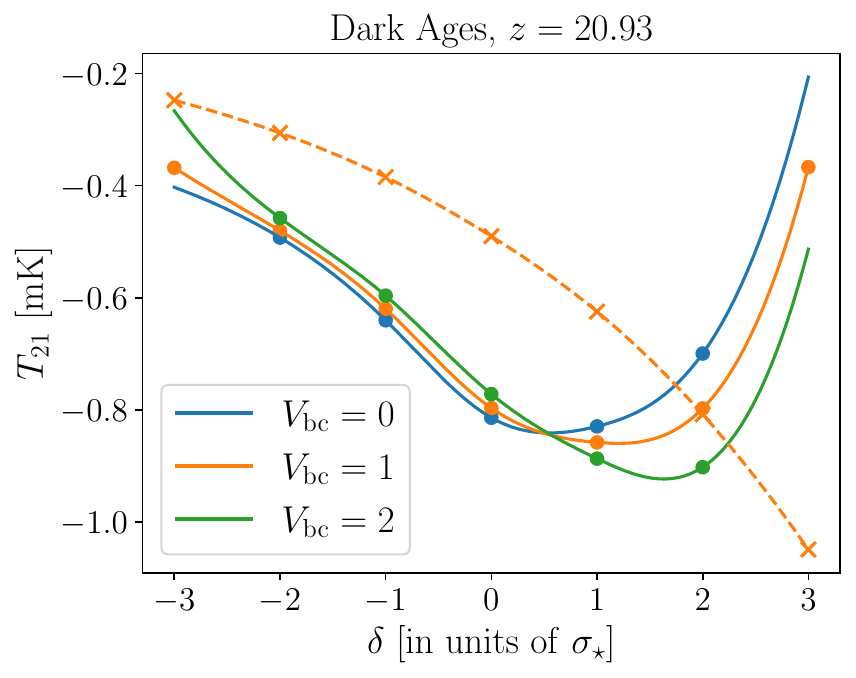}\hspace{-0.5em}
\includegraphics[width=0.37\textwidth]{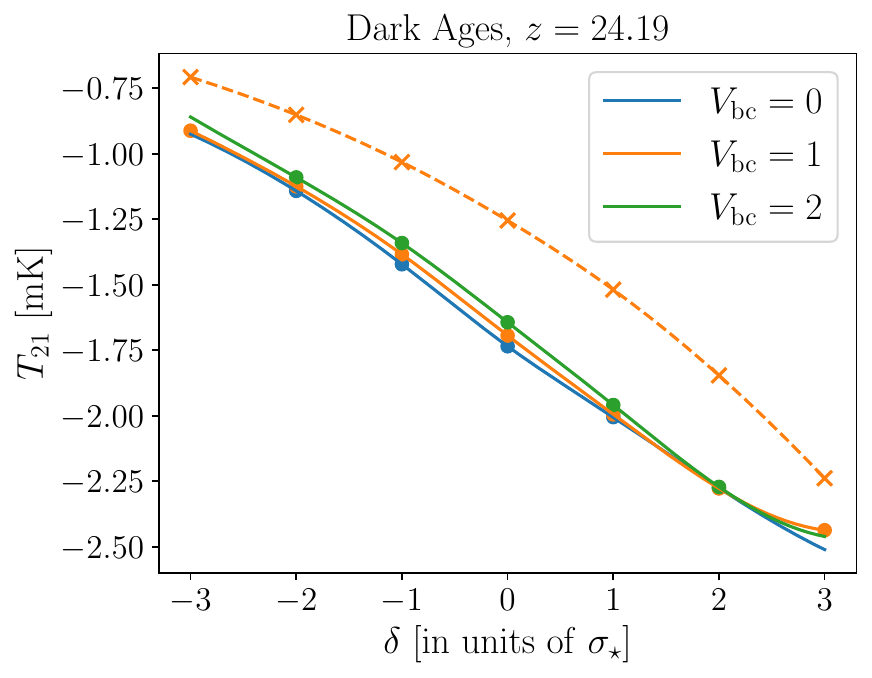}

\includegraphics[width=0.37\textwidth]{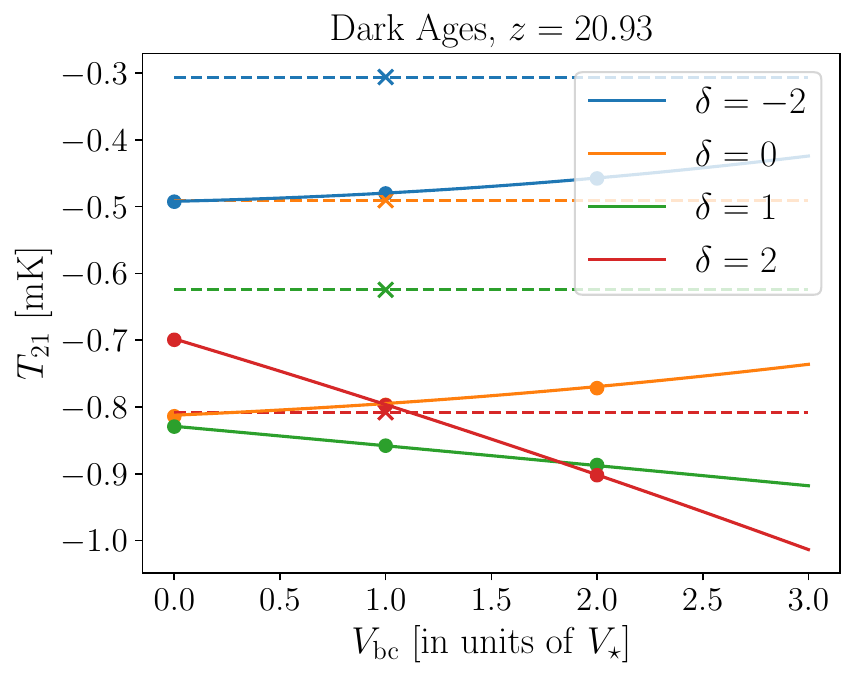}\hspace{-0.5em}
\includegraphics[width=0.37\textwidth]{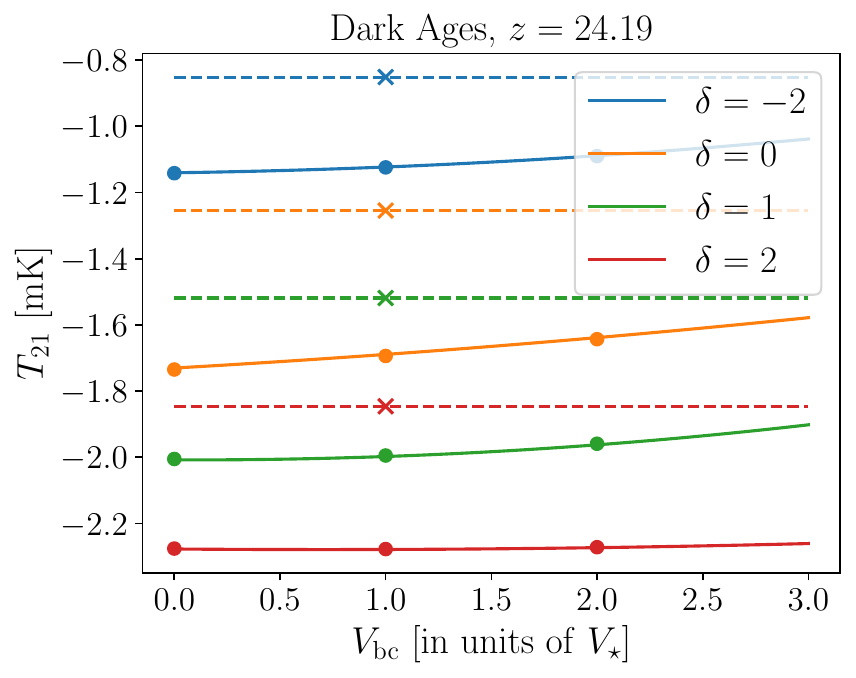}

\includegraphics[width=0.37\textwidth]{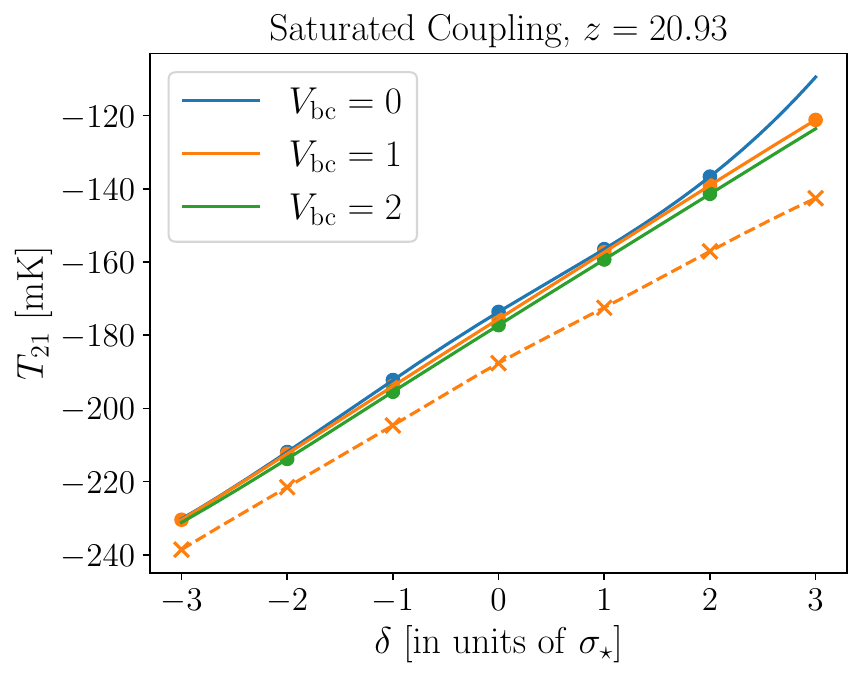}\hspace{-0.5em}
\includegraphics[width=0.37\textwidth]{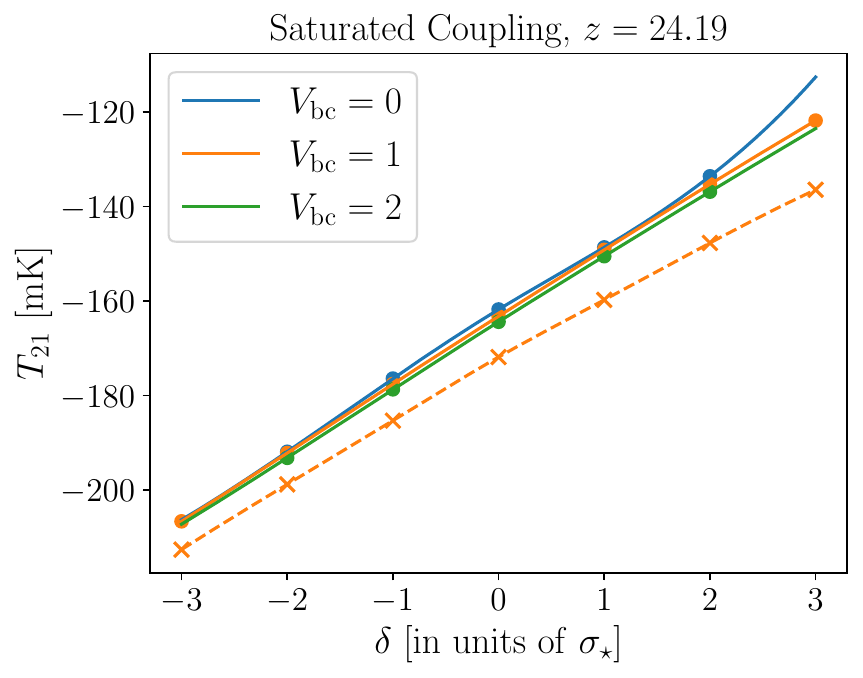}

\includegraphics[width=0.37\textwidth]{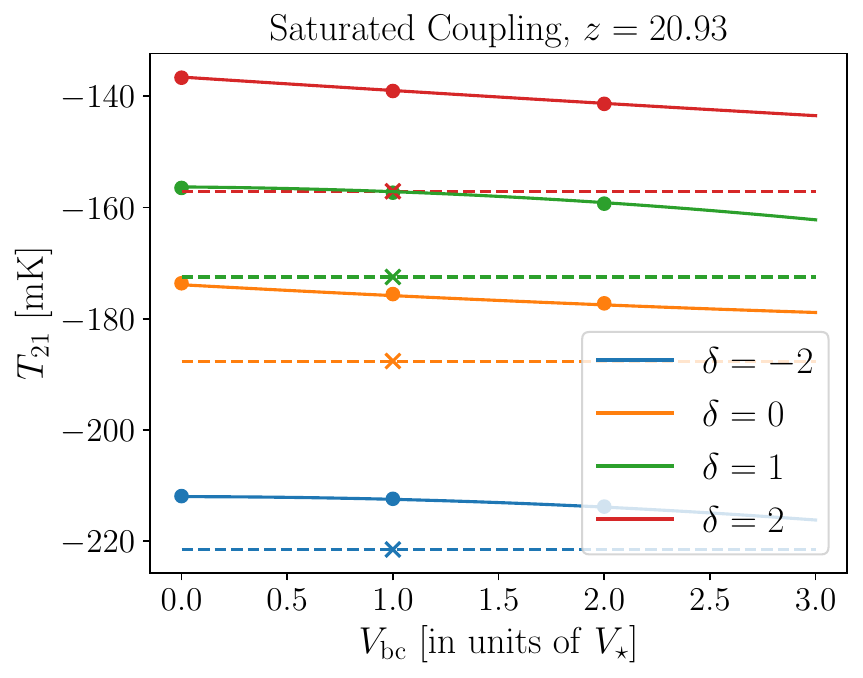}\hspace{-0.5em}
\includegraphics[width=0.37\textwidth]{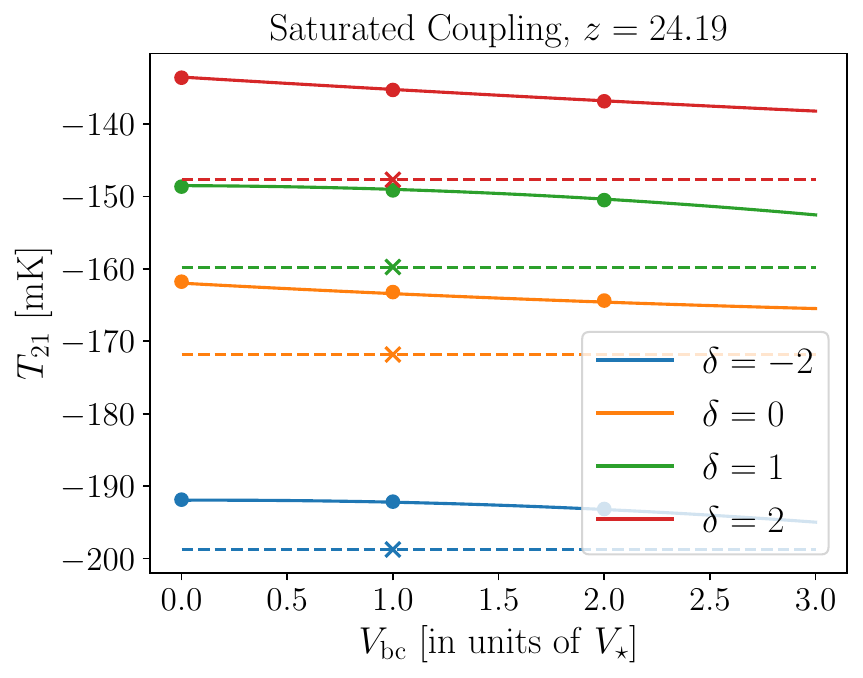}

\caption{Examples of the dependence of $T_{\rm{21}}$ on $\delta$ and $V_{\rm{bc}}$ (each in units of its cosmic standard deviation, see section~\ref{s:sim}). The points show the outputs of the hydrodynamical simulations, and the curves show our interpolation. We show the case with clumping (solid curves) or without, i.e., uniform simulation boxes without small-scale fluctuations (dashed curves); the latter case is independent of $V_{\rm{bc}}$ and the values are shown as $V_{\rm{bc}}=1$.}\label{fig:interp} 
\end{figure*}

In this Appendix we briefly illustrate our method of interpolating the 21-cm brightness temperature.
Fig.~\ref{fig:interp} shows the dependence of $T_{\rm{21}}$ on $\delta$ and $V_{\rm{bc}}$ (each in units of its cosmic standard deviation), at two example redshifts during cosmic dawn. We show two coupling cases, the Dark Ages and Saturated Coupling. We show the case with clumping (solid curves) or without (dashed curves). The uniform case (uniform simulation boxes without small-scale fluctuations) is independent of $V_{\rm{bc}}$ and the values are shown in that case as $V_{\rm{bc}}=1$. More generally, 
the dependence on $V_{\rm{bc}}$ is usually weak and nearly linear, and the dependence on $\delta$ is also nearly linear in many cases, and always smooth. For the interpolation we assumed a quadratic dependence on $V_{\rm{bc}}$ and applied a more general cubic spline interpolation for the dependence on $\delta$.

\bibliography{sample7}{}
\bibliographystyle{aasjournalv7}



\end{document}